\documentclass[10pt,conference,a4paper]{IEEEtran}
\usepackage{times,amsmath,epsfig}
  \usepackage[hyphens]{url}
  \usepackage{color}
  
  \newenvironment{noindlist2}
   {\begin{list}{\labelitemi}{\leftmargin=0.6em \itemindent=0em}}
   {\end{list}}

\title{Dynamic Configuration of Sensors Using Mobile Sensor Hub in Internet of Things Paradigm} 
\author{%
{Charith Perera{\small $~^{*\#1}$}, Prem Jayaraman{\small $~^{*2}$}, Arkady Zaslavsky{\small $~^{*3}$}, Peter Christen{\small $~^{\#4}$}, Dimitrios Georgakopoulos{\small $~^{*5}$} }
\vspace{1.6mm}\\
\fontsize{10}{10}\selectfont\itshape
$^{\#}$Research School of Computer Science, The Australian National University, Canberra, ACT 0200, Australia\\
\fontsize{9}{9}\selectfont\ttfamily\upshape
$^{4}$peter.christen@anu.edu.au

\fontsize{9}{9}\selectfont\ttfamily\upshape
\vspace{1.2mm}\\
\fontsize{10}{10}\selectfont\rmfamily\itshape
$^{*}$CSIRO ICT Center, Canberra, ACT 2601, Australia\\
\fontsize{9}{9}\selectfont\ttfamily\upshape
$^{1}$charith.perera@csiro.au,
$^{2}$prem.jayaraman@csiro.au,
$^{3}$arkady.zaslavsky@csiro.au,\\
$^{5}$dimitrios.georgakopoulos@csiro.au
}
\begin{document}
\maketitle
\begin{abstract} 
Internet of Things (IoT) envisions billions of sensors to be connected to the Internet. By deploying intelligent low-level computational devices such as mobile phones in-between sensors and cloud servers, we can reduce data communication with the use of intelligent processing such as fusing and filtering sensor data, which saves significant amount of energy. This is also ideal for real world sensor deployments where connecting sensors directly to a computer or to the Internet is not practical. Most of the leading IoT middleware solutions require manual and labour intensive tasks to be completed in order to connect a mobile phone to them. In this paper we present a mobile application called \textit{Mobile Sensor Hub} (\textit{MoSHub}). It allows variety of different sensors to be connected to a mobile phone and send the data to the cloud intelligently reducing network communication. Specifically, we explore techniques that allow MoSHub to be connected to cloud based IoT middleware solutions autonomously. For our experiments, we employed Global Sensor Network (GSN) middleware to implement and evaluate our approach. Such automated configuration reduces significant amount of manual labour that need to be performed by technical experts otherwise. We also evaluated different methods that can be used to automate the configuration process.
\end{abstract}

%
\section{Introduction}
\label{sec:Introduction}

\begin{figure}[h!]
 \centering
 \vspace{-0.33cm}
 \includegraphics[scale=.51]{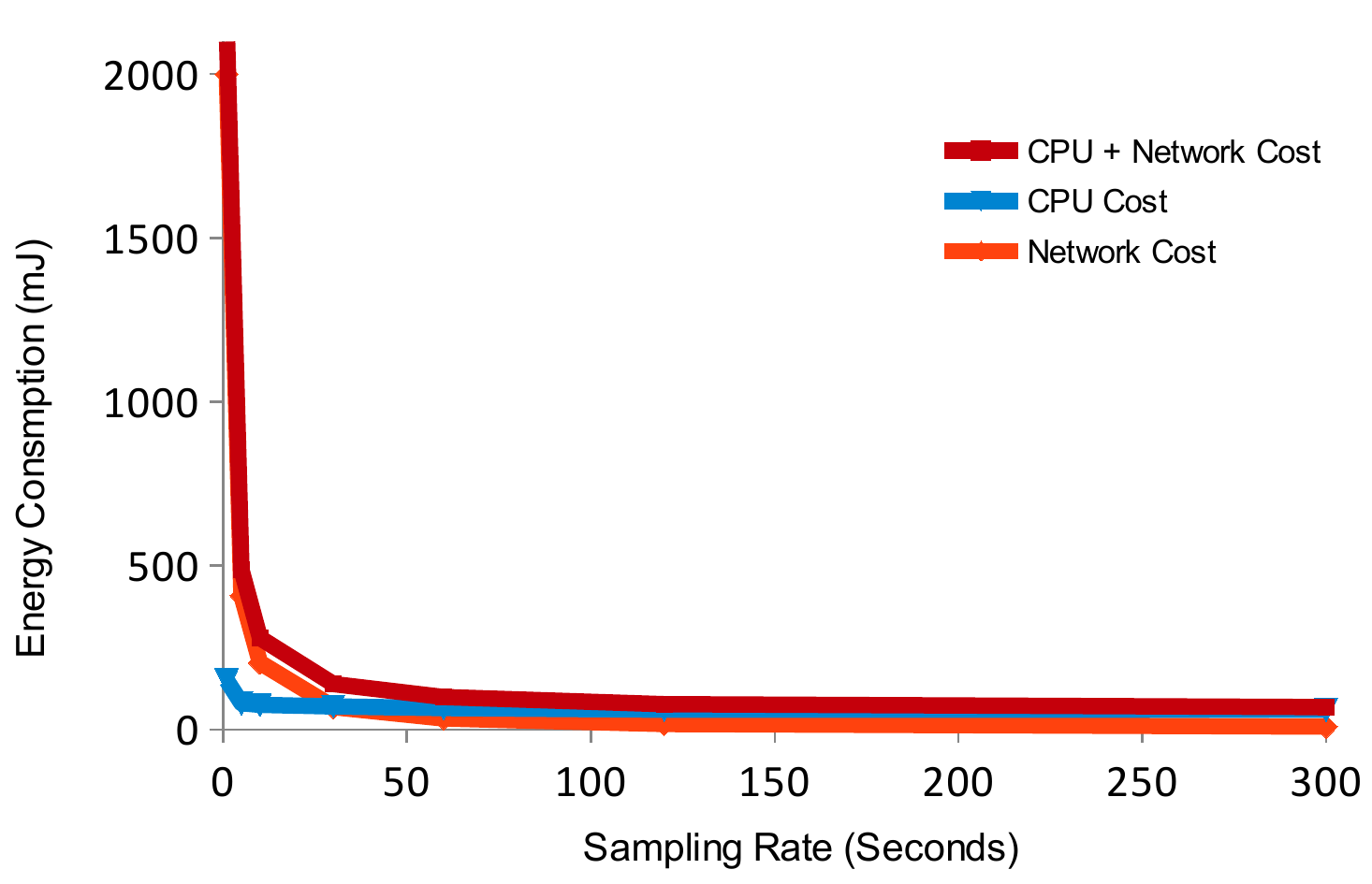}
\vspace{-0.33cm}	
 \caption{This graph shows how the energy consumption of Mobile Sensor Hub varies with the sampling rate. Only the network communication between server and MoSHub is considered.}
 
 \label{Figure:Energy_Consmption}	
\vspace{-0.40cm}	
\end{figure}

As we are moving towards the Internet of Things (IoT), the number of sensors deployed around the world is growing at a rapid pace. Market research has shown a significant growth of sensor deployments over the past decade and has predicted a significant increment of the growth rate in the future. Due to advances in sensor technology, sensors are getting
more powerful, cheaper and smaller in size, which has stimulated large scale deployments. Ultimately, these sensors will generate \textit{big data} \cite{ZMP003}. As shown in Figure \ref{Figure:Energy_Consmption}, communication of data from a sensor to a cloud application (or middleware) costs significant amount of energy in comparison to local data processing. Minimizing such communication using intelligent filtering and fusing techniques will save enormous amount of cost in IoT paradigm due to the magnitude.

\begin{figure}[h]
 \centering
    \vspace{-5pt}
 \includegraphics[scale=.55]{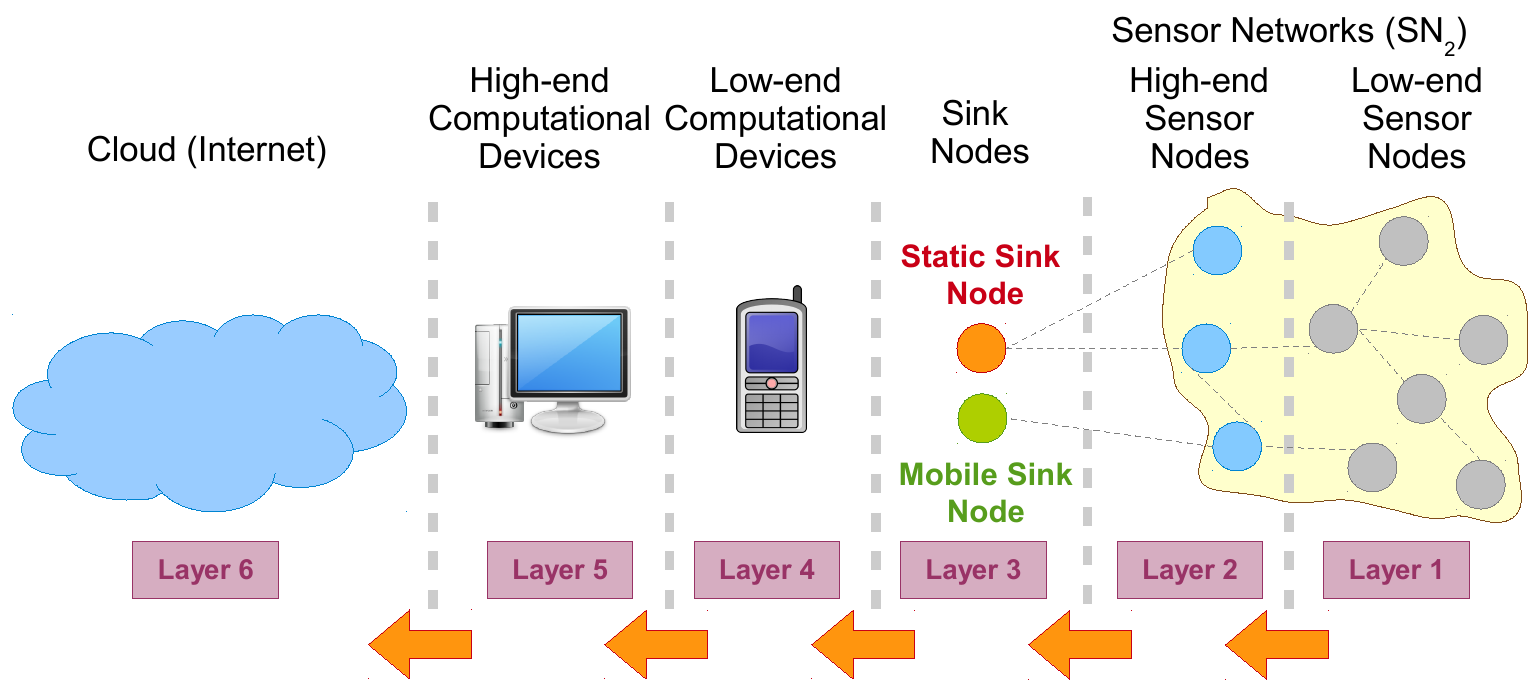}
 \caption{Layered structure of a sensor network: These layers are identified based on the capabilities posed by the devices. In IoT, this layered architecture may have additional number of sub layers as it is expected to comprises large variety  of sensing capabilities.}
 \label{Figure:Layered Structure on a Sensor Network}
   \vspace{-5pt}
\end{figure}

Typical structure of a sensor  network is presented in Figure  \ref{Figure:Layered Structure on a Sensor Network}. It comprises the most common components in a sensor network.  As we have shown, with the orange coloured arrows, data flows from right to left. Data is generated by the low-end sensor nodes  and high-end  sensor nodes. Then, data is collected by mobile and static sink nodes.  The sink nodes send the data to low-end computational devices. These devices perform a certain amount of processing on the sensor data. Then, the data is sent to high-end  computational  devices to be processed further. Finally,  data reaches the cloud where it will be shared, stored, and processed significantly.

Based on the capabilities of the devices involved in a sensor network, we have identified  six layers. Information can be processed in any layer. Capability means the processing, memory, communication,  and energy capacity. Capabilities increase  from layer one to layer six. Based  on our identification of  layers, it  is evident  that an ideal system should understand the capability  differences, and perform data management accordingly. For example, processing in the first three layers could reduce data communication.  However,  devices in the first three layers do not have a sufficient amount of energy  and processing  power to do comprehensive  data processing. Mobile phones have computational capabilities so it can fuse and filter data, which will help to reduce communication cost.

The rest of this paper is organized as follows: Section \ref{sec:Background_and_Related Work} describes the background and related work. The motivations and our contribution of this paper is presented in Section \ref{sec:Motivation_and_Our_Contribution}. In Section \ref{sec:Our_Approach}, we explain our proposed solution including MoSHub application in detail. Implementation and results of the evaluations are presented in Section \ref{sec:Implementation} and \ref{sec:Evaluation} respectively. Section \ref{sec:Lessons_Learned_and_Future_Work}  summarizes the lessons learned from this research.

\section{Background and Related Work}
\label{sec:Background_and_Related Work}

This work is based on two of our previous research efforts. In \cite{ZMP001}, we proposed a model called DAM4GSN that captures data using sensors built into the mobile phones. In this paper, we extend the support towards connecting external sensor. Further, we improved GSN middleware in such a way that it can dynamically generate custom wrappers\footnote{A program code (e.g. Java) that directly communicates with a sensor and retrieves data. It is a Java class that adheres to a specification.} for each MoSHub at run time instead of using a generic wrapper. In \cite{ZMP002}, we proposed ASCM4GSN architecture that automates the process of developing wrappers. We developed a XML based specification called Sensor Device Definition (SDD) that is capable of generating GSN wrappers. Focus of paper \cite{ZMP002} was to generate wrapper for sensors that uses manufacture released APIs. However, we utilized this technique in this work to generate customized wrappers for each MoSHub.

There are several other commercial solutions available: TWINE (supermechanical.com), Ninja Blocks (ninjablocks.com), and Smart Things (smartthings.com). All these solutions focus on event detection using If-THEN rules in smart environments. However, none of them support complex query processing capabilities similar to GSN. Their, automated configuration works only with limited number of devices they supports. Further, our plugin architecture allows to add more capabilities to MoSHub via existing app stores.


The Global Sensor Network (GSN) \cite{P022}, the IoT middleware we employed in this work, is a platform aimed at providing flexible middleware to address the challenges of sensor data integration and distributed query processing. It is a generic data stream processing engine. GSN has gone beyond the traditional sensor network research efforts such as routing, data aggregation, and energy optimisation. The design of GSN is based on four basic principles: simplicity, adaptivity, scalability, and light-weight implementation. GSN middleware simplifies the procedure of connecting heterogeneous sensor devices to applications. Specifically, GSN provides the capability to integrate, discover, combine, query, and filter sensor data through a declarative XML-based language and enables zero-programming deployment and management. Further, we are engaged in extending GSN middleware towards OpenIoT \cite{P377} by adding more capabilities. The above reasons lead us to choose GSN for our experiments. Our findings do not depend on any specific middleware and remain open to be used in any solution that needs mobile devices to be used as sensor hubs.

\section{Motivations and Our Contribution}
\label{sec:Motivation_and_Our_Contribution}

As depicted in Figure \ref{Figure:Layered Structure on a Sensor Network}, sensor data passes through different layers. In order to save network communication cost, we should be able to filter sensor data. Low computational devices such as mobile phones can be used to achieve above task. However, we need to understand what kind of operations can be done in mobile phones and what kind of operation need to be performed in server computers depending on their complexity and CPU cost. State-of-the-art mobile phones have many capabilities that make them ideal to be used in sensor data management in IoT domain. Mobile phones have built in wireless communication capabilities such as bluetooth, and WiFi. Further, these capabilities can be extended by connecting ZibBee modules via microUSB ports. Therefore, mobile phones can ideally be used as sensor data collecting devices. Technologies such as 3G and 4G allows transferring collected data to the cloud from place where WiFi networks are not available (e.g. environmental monitoring and agricultural domain). Latest mobile phones have up to 1.7 GHz Dual or Quad Krait CPU, 2 GB RAM and 8 GB internal storage. Therefore, mobile phones can ideally be used as sensor data processing devices as well.

The above-mentioned capabilities show that mobile phones can be  used as hubs. Mobile phones can offer the functionality of collect, process, and communicate sensor data to the cloud for further processing. The research challenge is \textit{how to connect mobile phones into an IoT middleware solution autonomously}. At a given time, variety  of different sensors may connected to a mobile phone locally. However, IoT middleware does not know about details of those sensors (what type of data or how much data to be expected from each mobile phone). In such a situation, \textit{how an IoT middleware can be configured  autonomously so it can accept data streams send by mobile phones or similar devices}. This is the research question we addressed in this paper.

\begin{figure}[h]
 \centering
 \vspace{-0.33cm}
 \includegraphics[scale=.35]{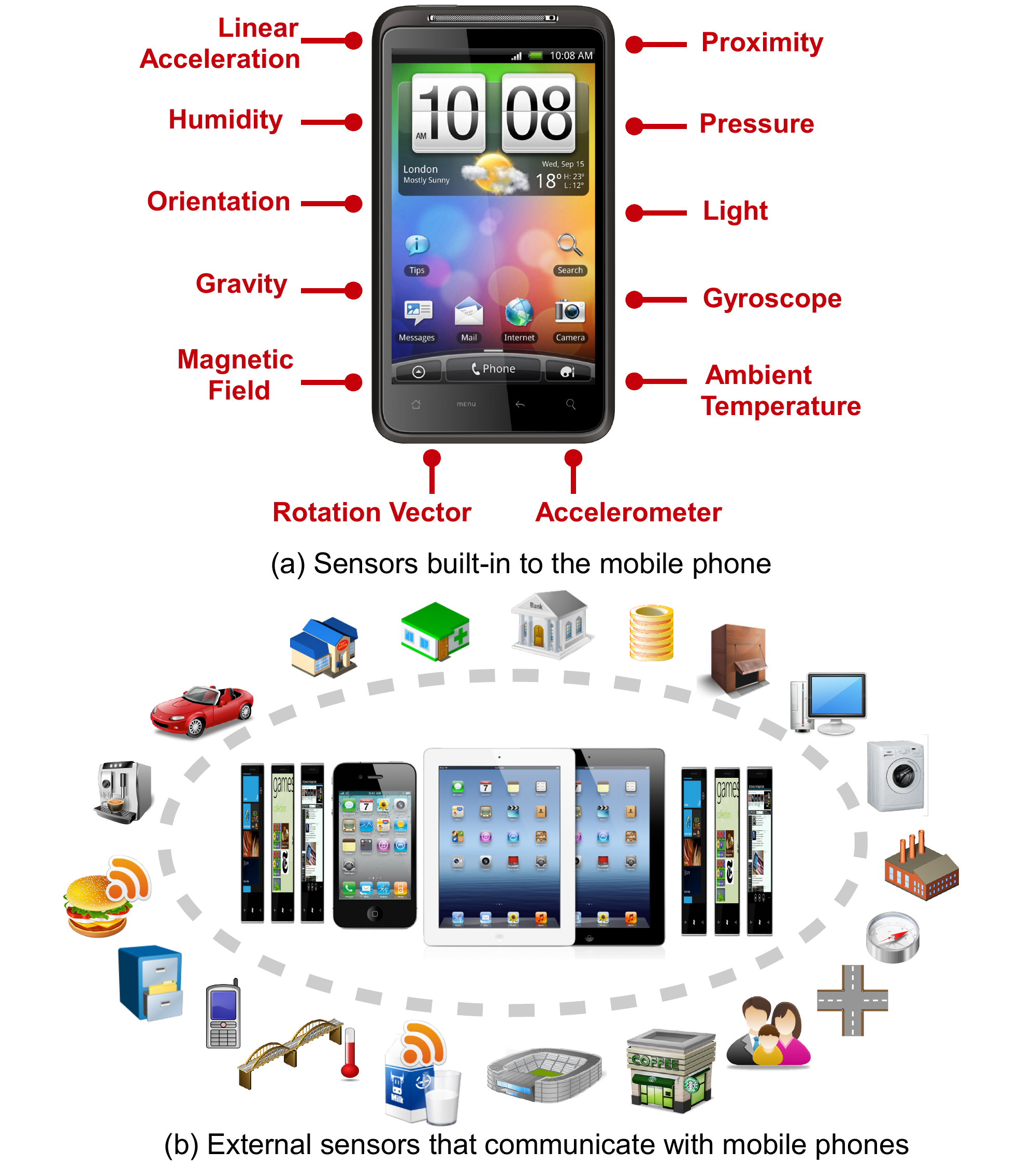}
\vspace{-0.33cm}	
 \caption{Sensors in Smart Environments}
 
 \label{Figure:MobilePhoneSensors}	
\vspace{-0.33cm}	
\end{figure}

The vision of the IoT is heavily  energised by statistics and predictions.  We present some statistics to justify our focus on the IoT and mobile computing and to show the magnitude of the challenges. It is estimated that there about 1.5 billion Internet-enabled PCs and over 1 billion Internet-enabled  mobile phones  today. These two categories  will  be joined with Internet-enabled devices (smart objects \cite{P041}))  in the future. By 2020, there will  be 50 to 100 billion devices connected to the Internet \cite{P029}.

In this paper, we propose a lightweight mobile application, called \textit{Mobile Sensor Hub} (MoSHub), which allows to connect and retrieve sensor data using wireless communication techniques such as WiFi and bluetooth from external sensors easily. We employed a plug-in architecture called Android Interface Definition Language (AIDL)\footnote{http://developer.Android.com/guide/components/aidl.html} provided in Android platform to facilitate plug and play configuration between external sensors and MoSHub application.

We propose a model that can configure the communication between mobile phone-based MoSHub and server-based GSN middleware.  This automated configuration reduces significant amount of manual labour need to be performed by technical experts otherwise. We extended and applied our previously proposed ASCM4GSN \cite{ZMP002} approach to generate programming code at runtime which enable dynamic configuration.

Finally, we compare two possible approaches that can be used to automate the configuration in many perspective as presented in Section \ref{sec:Implementation} in order to explore the most suitable approach to be used in MoSHub. We also carried out preliminary evaluations on scalability of the plug-in architecture.

\section{Our Approach}
\label{sec:Our_Approach}

In this section, we discuss our proposed solution in detail. First, we provide a high-level overview of our approach. Next, we explain how both client side and server side autonomous configuration works. Then, we describe the MoSHub application.  Throughout our discussion, we highlight possible alternative approaches and justifications on our choice. In Section \ref{sec:Implementation}, we evaluate and justify what approach is more appropriate based on experimentation results.

\subsection{High-level Overview of the System}
\label{sec:OA:High_level_Overview}

\begin{figure}[t]
 \centering
 \includegraphics[scale=.44]{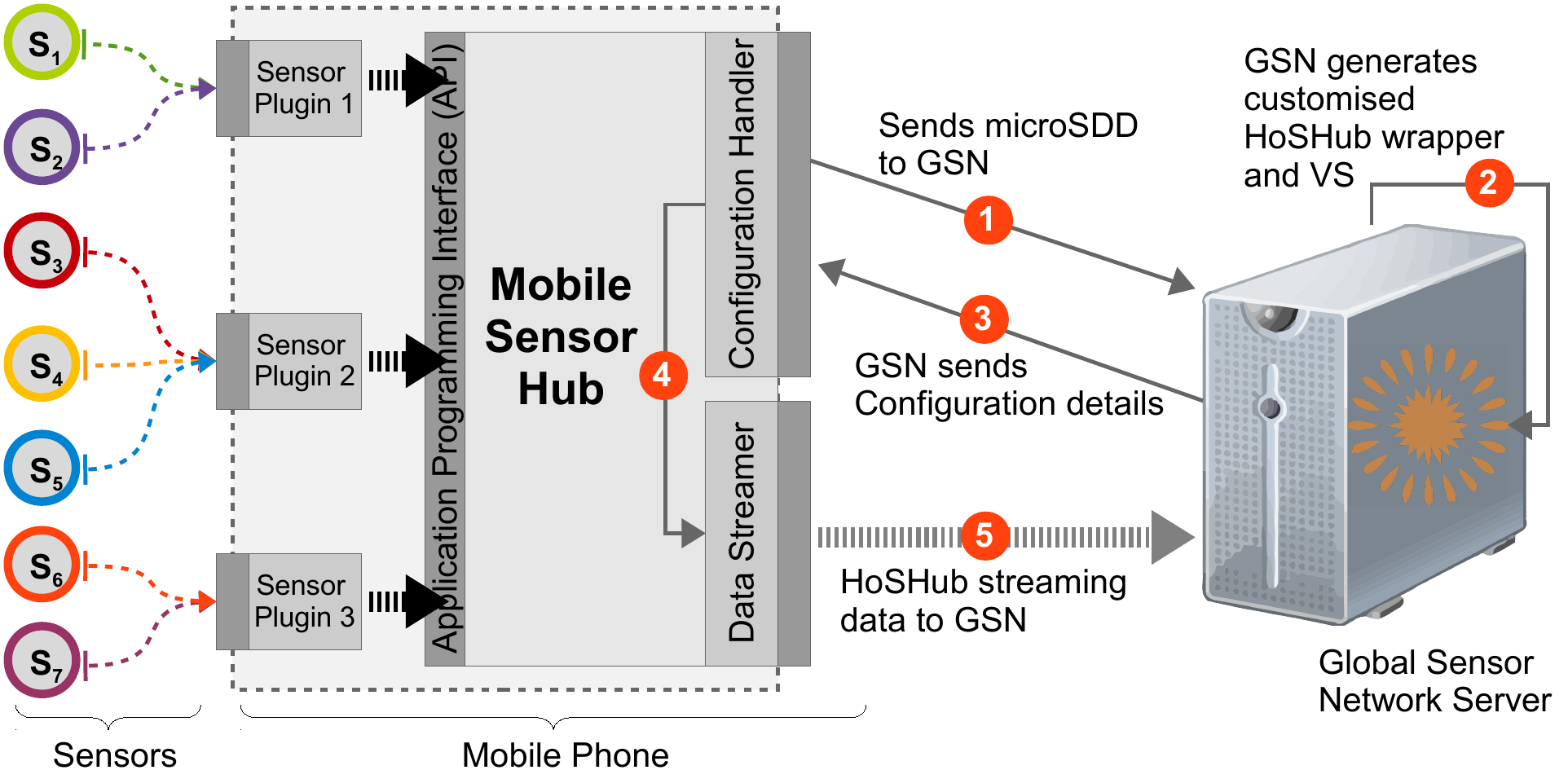}
\vspace{-0.13cm}	
 \caption{Main steps of the automated dynamic configuration process  that connects a MoSHub to a GSN server. Numbers show the order of execution.}
 
 \label{Figure:System_Overview}	
\vspace{-0.53cm}	
\end{figure}

\begin{figure}[h]
 \centering
 \vspace{-0.23cm}
 \includegraphics[scale=.98]{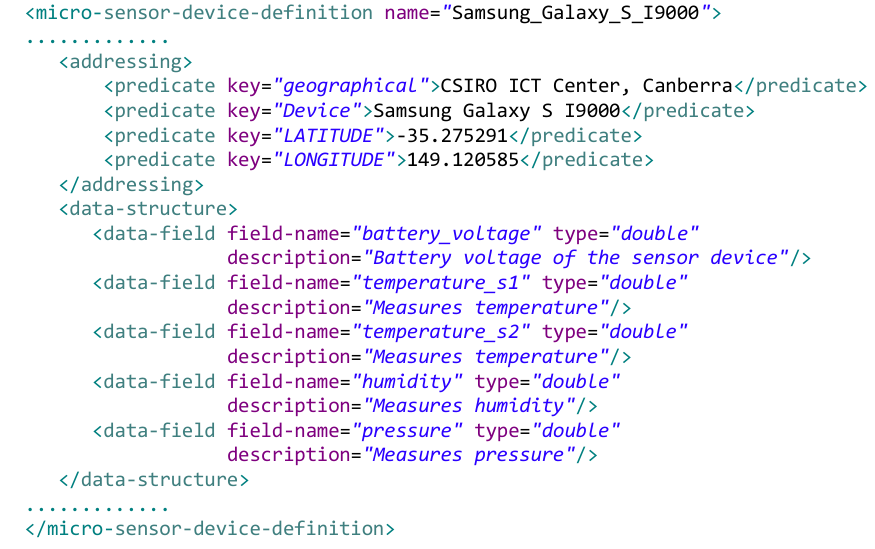}
\vspace{-0.23cm}	
 \caption{Sample $\mu$SDD  snippet which contains information about data produce by all the sensors connected to a MoSHub and context such a location.}
 
 \label{Figure:MicroVS}	
\vspace{-0.13cm}	
\end{figure}

The high level communication between MoSHub and GSN server is depicted in Figure \ref{Figure:System_Overview}. We can explain how automated configuration works in order of activities as follows. MoSHub is an application that need to be installed in an Android mobile phone, which is intended collect data from both the internal and external sensors. Even though we use the term mobile phone, in actual architecture what we need is some device that has the capabilities, similar to a mobile phone, such as WiFi, bluetooth, CPU, memory. In the future, we expect there would be devices, powered by Android, specifically design for IoT paradigm. Such environment will add more value to our research and open up more opportunities. Different types of sensors can be connected to MoSHub via different wireless technologies. Then, MoSHub generates a micro sensor device definition ($\mu$SDD) file based on sensors connected to it. $\mu$SDD is different from GSN virtual sensor definition (VS) and it is somewhat similar to SDD \cite{ZMP002}. Figure \ref{Figure:MicroVS} shows a $\mu$SDD definition file snippet.

The reason for generating a $\mu$SDD without directly generating a SDD is two fold. First, though both definitions share some amount of similarities, they should be able to extended independently from the each other depending on the requirements arises in the future. Second reason is the network communication. We want keep the packet size to the minimum, which will save energy that it take to generate the file as well as in network communication. Further, keeping only the minimum amount of information that is specific to each situation makes it easier and faster to process.

MoSHub sends the $\mu$SDD to the GSN server. GSN server then process the $\mu$SDD and generates a GSN wrapper class file that is specific to each individual MoSHub. GSN automatically compile the newly generated class file and add it to the wrapper repository. However, before generating a new class file GSN search for an existing matching class. In such case, GSN will use that class instead of generating a new one. The process of generating a wrapper based on a given SDD specification is described in our previous work \cite{ZMP002}. In that perspective, $\mu$SDD acts same as the SDD.

\begin{figure}[h]
 \centering
 \includegraphics[scale=.80]{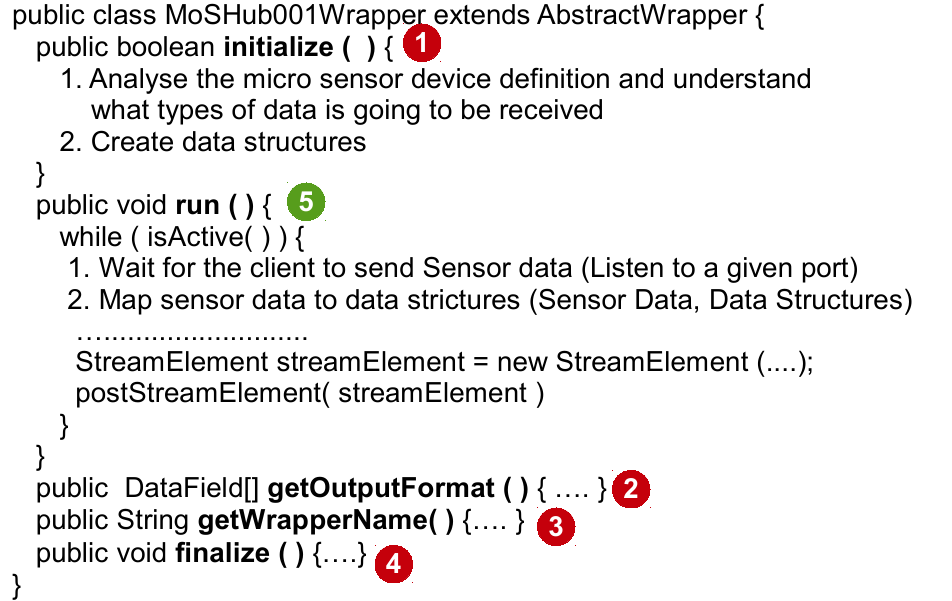}
\vspace{-0.33cm}	
 \caption{The Structure of a Typical MoSHub Wrapper}
 
 \label{Figure:GSNWrapper}	
\vspace{-0.33cm}	
\end{figure}

Figure \ref{Figure:GSNWrapper} depicts the structure of a GSN wrapper class. The content of the class would be generated based on the information provided in $\mu$SDD. Explanations are provided in \cite{ZMP001}. There are five methods in a typical GSN wrapper class. Method (1) runs only once and method (2) to (4) may run occasionally. In contrast, method (5)  will run every time when a new data stream receives.

We tested another approach that can  be used to achieve above functionality. Without creating customize wrapper for each MoSHub, we developed a generic MoSHub wrapper that can retrieve any amount of sensor data. This generic MoSHub wrapper can configure its internal data structures depending on how many data items are sent by each MoHub. However, during our performance evaluation, it was found that using generic wrapper is inefficient compared to generating customised wrapper for each sensor. Details are presented in Section \ref{sec:Implementation} and \ref{sec:Lessons_Learned_and_Future_Work}. Another ongoing study, we are conducting focusing on adding context discovery functionality to GSN middleware, also showed that generating customized wrapper for each MoSHub approach is better in term of  extensibility.

\begin{figure}[h]
 \centering
 \includegraphics[scale=.93]{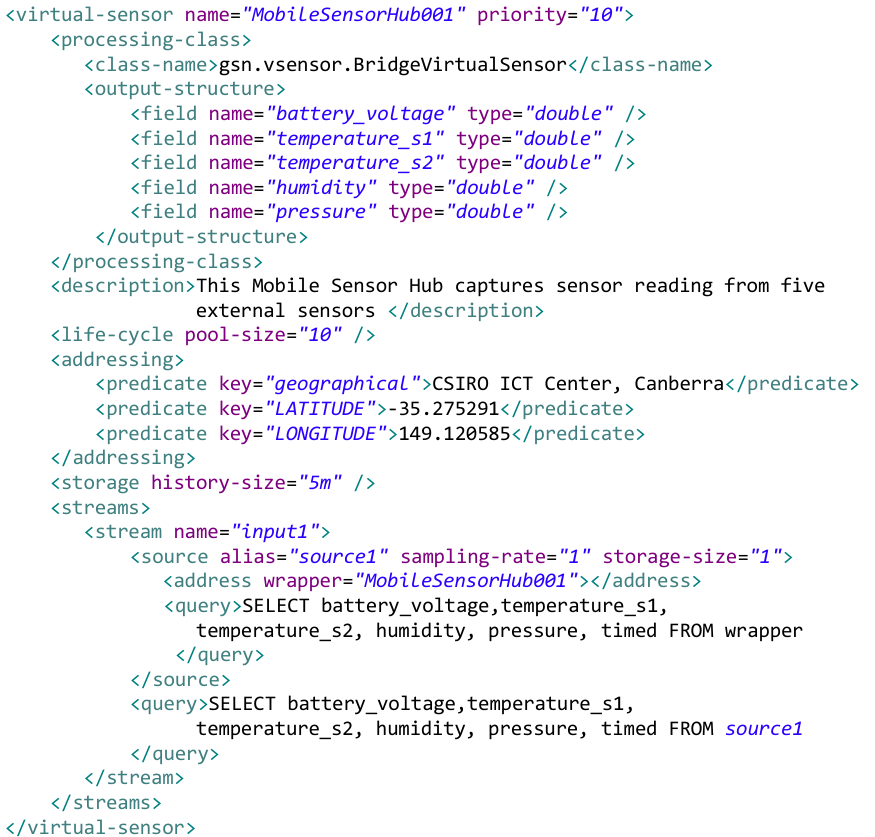}
\vspace{-0.23cm}	
 \caption{Virtual sensor definition (VSD) generated by GSN middleware during the automated configuration process using $\mu$SDD which sends by MoSHub.}
 
 \label{Figure:XML}	
\vspace{-0.33cm}	
\end{figure}

\begin{figure}[h]
 \centering
 \includegraphics[scale=.97]{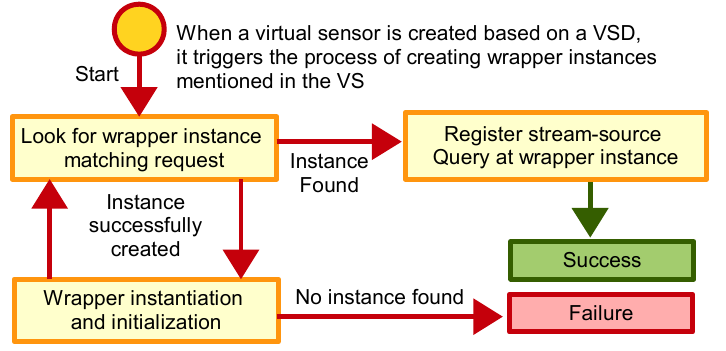}
\vspace{-0.13cm}	
 \caption{Wrapper life cycle of the GSN middleware}
 
 \label{Figure:Wrapper_Life_Cycle}	
\vspace{-0.53cm}	
\end{figure}

After generating MoSHub wrapper, GSN generates a virtual sensor definition (VSD) using the information provided in the $\mu$SDD. GSN VSD is explained in details in \cite{P022}. Even though a virtual sensor definition can combine data coming from multiple wrappers, in default automated configuration process, GSN creates a dedicated virtual sensor definition for each MoSHub wrapper. Figure \ref{Figure:XML} presents a sample MoSHub virtual sensor definition. When GSN generates a VSD file, it triggers the virtual sensor creation processes. This process triggers the specified wrapper to be created. The wrapper that correspond to each stream source is defined under the address element in the VSD file. This process sends a Wrapper Connection Request (WCR) to the wrapper repository in the GSN server. WCR is an object, which contains a wrapper name and its initialisation parameters as defined in the virtual sensor definition. Whenever a WCR is generated at the virtual sensor loader, it will be sent to the wrapper repository. Then, steps are followed as depicted in Figure \ref{Figure:Wrapper_Life_Cycle}. A detailed description of this process is presented in \cite{ZMP001}.

Once the wrapper instance is ready to receive data, GSN sends configuration detail to the configuration handler of the MoSHub application. This information contains the port number where MoSHub needs to send data. At this point, automated configuration process completes. Finally, MoSHub starts streaming data to the GSN. When a new sensor connects to MoSHub or existing sensor disconnects from MoSHub, the automated configuration process need to be executed again.

\subsection{Mobile Sensor Hub (MoSHub)}
\label{sec:OA:Mobile_Sensor_Hub}

MoSHub is a mobile application that collects, combines, processes, and sends sensor data to a GSN server. Communication between external sensors and MoSHub is conducted through independent software layer called \textit{plug-ins}. MoSHub provides a specification that defines how developers should develop plug-ins that will be able to communicate with MoSHub application. Due to space limitation, we do not describe those specifications in this paper. In brief, the specification guides the developers on how to name their plug-ins, packages, and provides an interface (including list of methods need to be implemented, common data structures and so on) as an aidl file. The operations that can be conducted by a given plug-in is limited only by developers capability and Android platform. As long as plug-ins are adhered to the provided specification, they will be able to communicate with MoSHub application.

In order to generate $\mu$SDD file, MoSHub communicates with every active plug-in that collects data from a external or internal sensor. Each plug-in should at least provide the category/name of the sensor they are communicating with (e.g. temperature\_s1) and type of data that connected sensor generates (e.g. int, double, string). Once MoSHub gathers minimum amount of information from all the plug-ins, it generates the $\mu$SDD file. It may also include available context information such as location.

\section{Implementation}
\label{sec:Implementation}

\begin{figure}[h!]
 \centering
 \includegraphics[scale=.80]{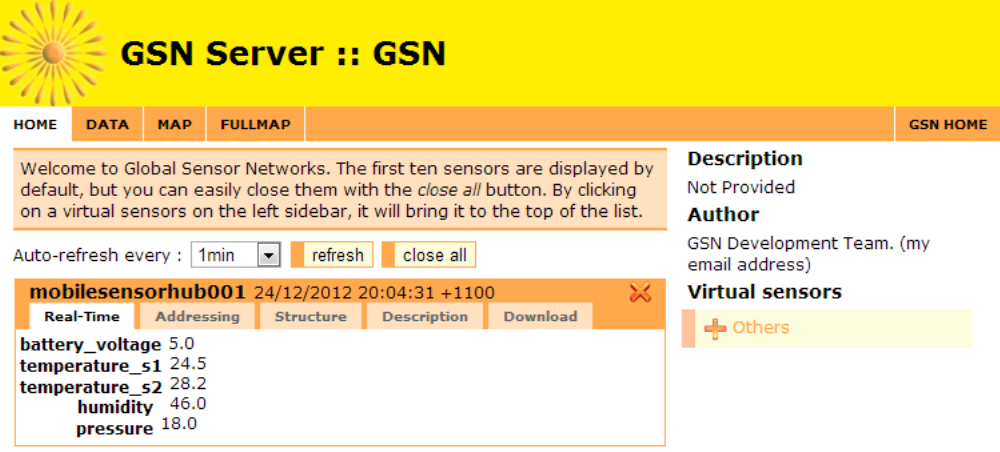}
\vspace{-0.30cm}	
 \caption{This is a sample web based user interface of the GSN middleware that shows a MoSHubs is connected to it.}
 
 \label{Figure:GSN_UI}	
\vspace{-0.3cm}	
\end{figure}

\begin{figure}[h]
 \centering
 \includegraphics[scale=.35]{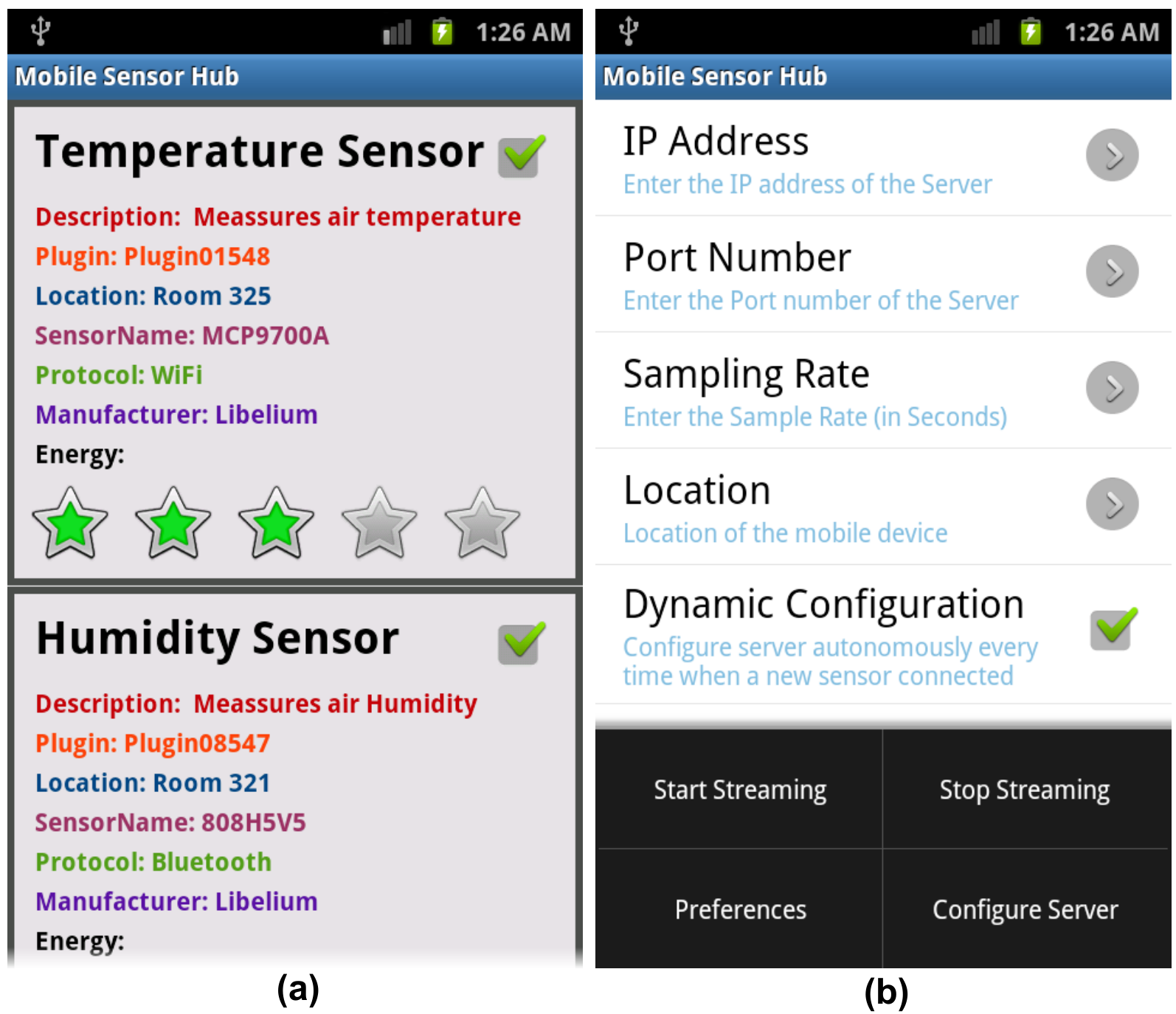}
\vspace{-0.33cm}	
 \caption{These are main user interfaces of the MoSHup application. (a) shows list of sensors that are connected to MoSHup though different plug-ins. (b) shows configuration screen of the MoSHup application.}
 
 \label{Figure:MoSHub_User Interfaces}	
\vspace{-0.23cm}	
\end{figure}

We conducted all evaluations and experiments using a Samsung Galaxy S GT-I9000 mobile phone, which runs Android platform 2.3.6. GSN middleware was installed on a laptop with Intel Core i5 CPU and 4GB RAM. Network communications are conducted through CSIRO ICT centre WiFi network. Figure \ref{Figure:GSN_UI} shows a web interface of GSN middleware and Figure \ref{Figure:MoSHub_User Interfaces} shows main user screens of the MoSHub application. 


\section{Evaluation}
\label{sec:Evaluation}

\begin{figure}[h]
\vspace{-0.33cm}
 \centering
 \includegraphics[scale=.46]{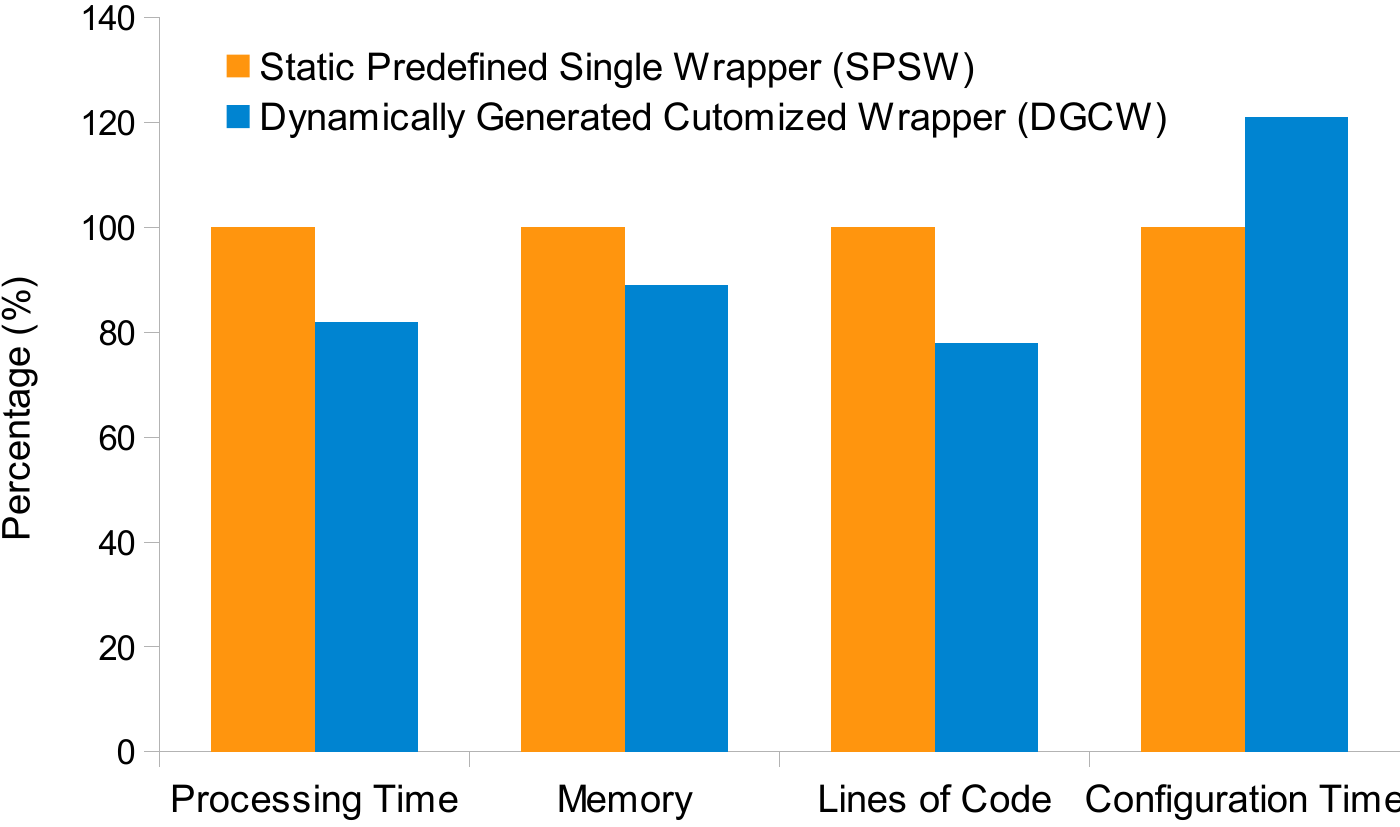}
\vspace{-0.33cm}	
 \caption{Comparison of SPSW and DGCW Approaches}
 
 \label{Figure:Comparison}	
\vspace{-0.53cm}	
\end{figure}

\begin{figure}[h]
\vspace{-0.13cm}
 \centering
 \includegraphics[scale=.46]{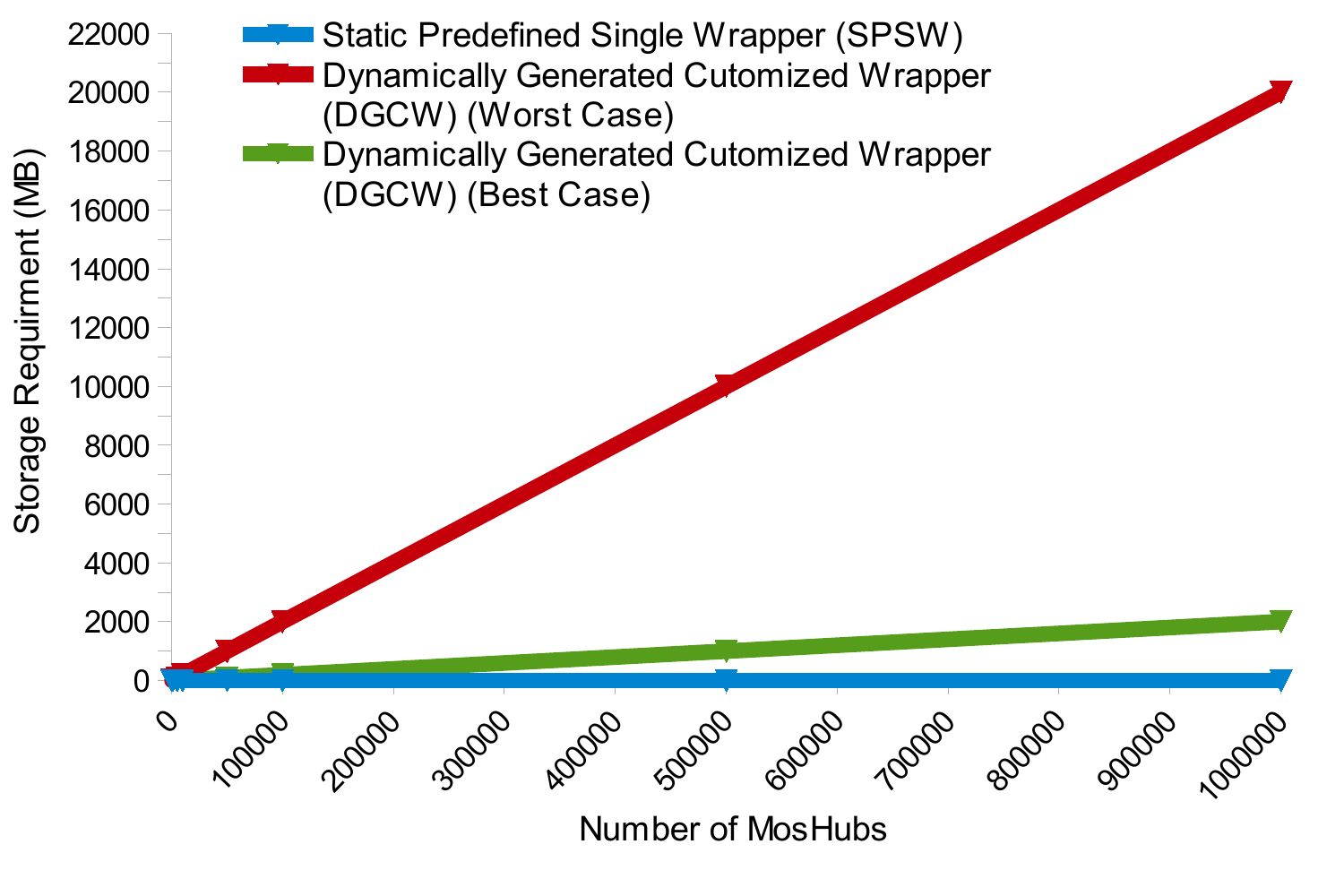}
\vspace{-0.53cm}	
 \caption{Storage Requirments of SPSW and DGCW (Estimated)}
 
 \label{Figure:Storage_Requirments}	
\vspace{-0.33cm}	
\end{figure}

\begin{figure}[h]
 \centering
 \vspace{-0.33cm}
 \includegraphics[scale=.46]{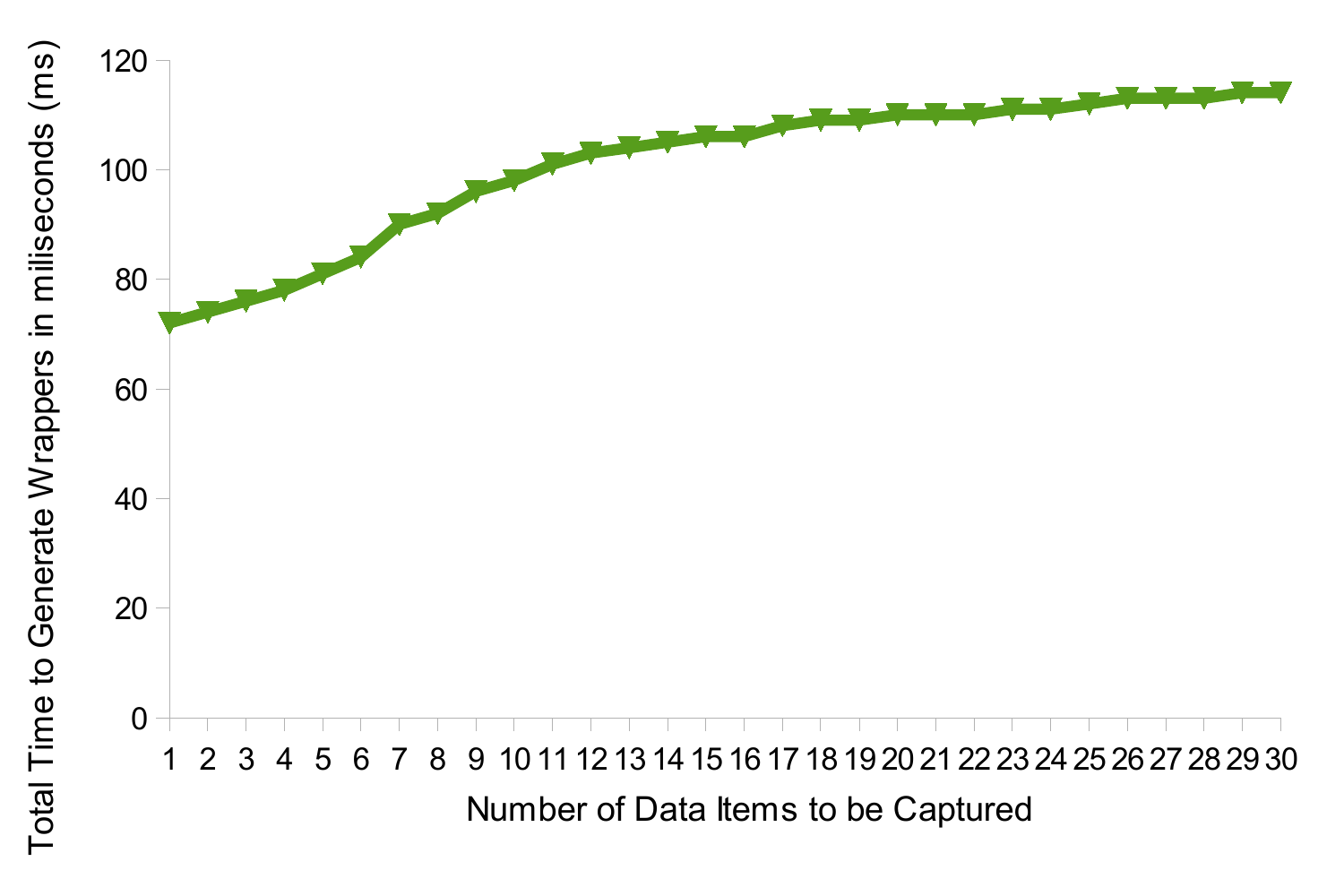}
\vspace{-0.53cm}	
 \caption{Performance Measurement of GSN Wrapper Generation}
 
 \label{Figure:Code_Generation}	
\vspace{-0.33cm}	
\end{figure}

In this section, we present results of several experiments we conducted in order to evaluate the performance and suitability of the approaches we proposed. Figure \ref{Figure:Comparison} graph shows a comparison of two approaches we explained earlier, in Section \ref{sec:Our_Approach}, in four different perspectives. Four comparisons are conducted using different measurement units: processing time (in milliseconds), memory (KB), lines of code (number of lines), and automated configuration time (in milliseconds). A MoSHub with eight sensors connected to it used for evaluations. In order to combine all perspectives into one graphs, we converted all of them to percentages. Static predefined single wrapper (SPSW) approach is kept as 100\%. Dynamically generated customized wrappers (DGCW) approach is graphed in compared to SPSW. Therefore, this graph show how much DGCW approach is efficient or inefficient compared to SPSW as a percentage. For example, DGCW takes 18\% less processing time than SPSW. Figure \ref{Figure:Storage_Requirments} shows how storage requirement of the GSN middleware varies when number of MoSHubs connected to GSN increases in two different approaches. Figure \ref{Figure:Code_Generation} shows how much time it takes to generate a wrapper based on micro sensor device definitions ($\mu$SDD) when complexity increases. Interpretation of these results are presented in Section \ref{sec:Lessons_Learned_and_Future_Work}.

\section{Lessons Learned and Future Work}
\label{sec:Lessons_Learned_and_Future_Work}

\begin{figure}[t]
 \centering
 \includegraphics[scale=.46]{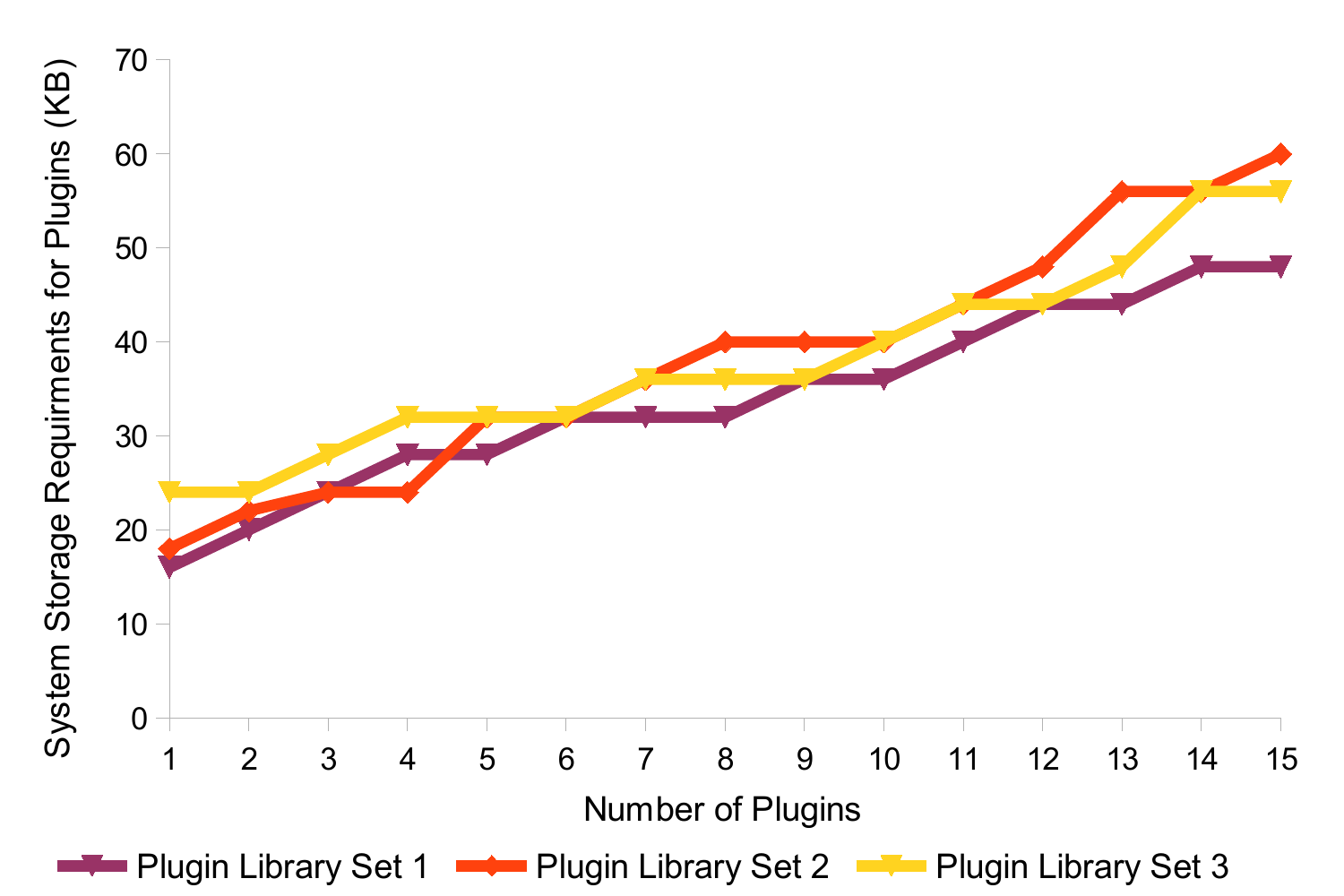}
\vspace{-0.33cm}	
 \caption{This graph shows the amount of memory (system storage of the mobile phone) required by plug-ins when number of plug-ins increases. The storage is measured in Kilobytes (KB). We developed three sets of plug-in libraries where each contains 15 plug-ins. Each library comprises of different plug-ins that are capable of retrieving data from different wasp sensors from libelium (www.libelium.com) using WiFi and bluetooth.}
 \vspace{-0.50cm}
 \label{Figure:Plugin_Storage}	
	
\end{figure}

Our ultimate objective is to develop a mobile application that can be used to collect data from external sensors, intelligently process, and communicate them to servers over the Internet. In order to achieve above objective, both client (i.e. mobile application) and server (i.e. IoT middleware) should be able to configure themselves autonomously so they can work together efficiently. In this work, we explored approaches that can be used to automate the configuration process. Further, we employed plug-in architecture to increase the support towards external sensors. Lessons we leant can be listed as follows:

\begin{noindlist2}

\item As depicted in Figure \ref{Figure:Comparison}, DGCW can reduce data stream processing time up to 18\%. Therefore, generating customized wrapper for each MoSHub based on the sensors (e.g. number of sensors and type of sensors) connected to them, is more efficient than using a generic wrapper that dynamically changes its data structures at runtime. In SPSW, all the exact details about data structure need to be understand via XML descriptions and then dynamically initialize during runtime (i.e. during wrapper initialization phase).

\item When consider memory requirements, 11\% can be saved by following DGCW approach. Further, wrapper in DGCW approach uses up to 22\% less lines of code compared to wrapper in SPSW depending on the class complexity (This is true until the number of data items need to be captured stays below 25). In contrast, DGCW approach take more time to configure each MoSHub to GSN middleware. However, each MoSHub will need to configure itself with GSN only when number of sensors connected to it changes. This will not happen regularly. Therefore,
DGCW approach is more efficient when all four factor considered together.

\item In DGCW approach, GSN needs to generate a customised wrapper code for each MoSHub. A typical wrapper is around 15-25KB in size. In SPSW approach, GSN needs only one wrapper. If we consider storage requirement factor in isolate manner, it seems SPSW is more efficient as depicted in Figure \ref{Figure:Storage_Requirments}. However, due to advances in computer hardware, storage is much cheaper than processing. For example, we can store one million different wrappers in a 20GB storage space. Therefore, when we take runtime efficiency into account, higher storage requirement of DGCW can be neglected. Further, GSN loads only one wrapper for each MoSHub to the memory. Therefore, no additional memory will be used in DGCW approach.

\item In SPSW approach, there is no requirement to generate wrapper  code at runtime. So there is no delay in configuration process. In contrast, DGCW approach needs to generate a wrapper code every time when MoSHub needs to be configured with GSN. (Note: This is only required when a MoSHub connects to GSN with specific sensor configuration for the first time. If a MoSHub connects to a GSN instance with same configuration for the second time, GSN will automatically select the previously generated wrapper code without creating a new wrapper). DGCW approach takes 70ms-120ms to create a wrapper code based on the complexity as depicted in Figure \ref{Figure:Code_Generation}.

\item As DGCW approach generates a customized wrapper for each MoSHub, it creates significant amount of opportunities. For example, DGCW approach allows adding context discovery functionality to GSN in the future, which is difficult to accomplish using SPSW approach.

\item According to Figure \ref{Figure:Plugin_Storage}, plug-in architecture seems promising in term of memory requirements. A plug-in library that comprises 15 different plug-ins needs only 40-60KB storage space. Therefore, plug-in architecture is scalable and suitable to be used in MoSHub. Based on our preliminary investigation, storage requirement for plug-ins is linear. However, there is an initial storage requirement of 20KB for meta-data and configuration information required by Android application model. Therefore, it is ideal to combine multiple plugins into libraries to minimize meta-data overhead.

\item Our preliminary investigation showed us that the issue of re-configuration MoSHub, due to local sensor connectivity changes, can be minimized by accepting null value over some period. The sensors, which disconnect from MoSHub due to technical failures will establish its connection back within limited time so we can avoid triggering costly re-configuration process.

\end{noindlist2}

\vspace{-0.06cm}

In our future work, we will evaluate plug-in architecture in a comprehensive manner in order to identify the scalability of our approach in term of energy and memory consumption at runtime. We will extend our evaluation platform towards different computational devices such as tablets with different hardware specifications. Further, we will explore techniques of using different protocols  such as bluetooth, WiFi, and ZigBee to detect sensors and then select appropriate plug-ins autonomously which will help to retrieve data from detected sensors using cloud repositories such as Android market.

\bibliographystyle{IEEEtran}

\bibliography{Bibliography}

\begin{thebibliography}{1}
\providecommand{\url}[1]{#1}
\csname url@samestyle\endcsname
\providecommand{\newblock}{\relax}
\providecommand{\bibinfo}[2]{#2}
\providecommand{\BIBentrySTDinterwordspacing}{\spaceskip=0pt\relax}
\providecommand{\BIBentryALTinterwordstretchfactor}{4}
\providecommand{\BIBentryALTinterwordspacing}{\spaceskip=\fontdimen2\font plus
\BIBentryALTinterwordstretchfactor\fontdimen3\font minus
  \fontdimen4\font\relax}
\providecommand{\BIBforeignlanguage}[2]{{%
\expandafter\ifx\csname l@#1\endcsname\relax
\typeout{** WARNING: IEEEtran.bst: No hyphenation pattern has been}%
\typeout{** loaded for the language `#1'. Using the pattern for}%
\typeout{** the default language instead.}%
\else
\language=\csname l@#1\endcsname
\fi
#2}}
\providecommand{\BIBdecl}{\relax}
\BIBdecl

\bibitem{ZMP003}
A.~Zaslavsky, C.~Perera, and D.~Georgakopoulos, ``Sensing as a service and big
  data,'' in \emph{International Conference on Advances in Cloud Computing
  (ACC-2012)}, Bangalore, India, July 2012.

\bibitem{ZMP001}
C.~Perera, A.~Zaslavsky, P.~Christen, A.~Salehi, and D.~Georgakopoulos,
  ``Capturing sensor data from mobile phones using global sensor network
  middleware,'' in \emph{IEEE International Workshop on Internet-of-Things
  Communications and Networking 2012 (PIMRC 2012-Workshop-IoT-CN12)}, Sydney,
  Australia, September 2012.

\bibitem{ZMP002}
------, ``Connecting mobile things to global sensor network middleware using
  system-generated wrappers,'' in \emph{International ACM Workshop on Data
  Engineering for Wireless and Mobile Access 2012 (ACM SIGMOD/PODS
  2012-Workshop-MobiDE)}, Scottsdale, Arizona, USA, May 2012.

\bibitem{P022}
\BIBentryALTinterwordspacing
K.~Aberer, M.~Hauswirth, and A.~Salehi, ``Infrastructure for data processing in
  large-scale interconnected sensor networks,'' in \emph{International
  Conference on Mobile Data Management}, May 2007, pp. 198--205. [Online].
  Available: \url{http://dx.doi.org/10.1109/MDM.2007.36}
\BIBentrySTDinterwordspacing

\bibitem{P377}
{OpenIoT Consortium}, ``Open source solution for the internet of things into
  the cloud,'' January 2012, \url{http://www.openiot.eu} [Accessed on:
  2012-04-08].

\bibitem{P041}
\BIBentryALTinterwordspacing
G.~Kortuem, F.~Kawsar, D.~Fitton, and V.~Sundramoorthy, ``Smart objects as
  building blocks for the internet of things,'' \emph{Internet Computing,
  IEEE}, vol.~14, no.~1, pp. 44 --51, jan.-feb. 2010. [Online]. Available:
  \url{http://dx.doi.org/10.1109/MIC.2009.143}
\BIBentrySTDinterwordspacing

\bibitem{P029}
H.~Sundmaeker, P.~Guillemin, P.~Friess, and S.~Woelffle, ``Vision and
  challenges for realising the internet of things,'' European Commission
  Information Society and Media, Tech. Rep., March 2010,
  \url{http://www.internet-of-things-research.eu/pdf/IoT_Clusterbook_March_2010.pdf}
  [Accessed on: 2011-10-10].

\end{thebibliography}

\end{document}